\documentclass{epl}

\usepackage{graphicx,float}
\usepackage{amssymb}
\usepackage[centertags]{amsmath}
\usepackage[dvips]{color}

\title{Scaling Behaviour of the Maximal Growth Rate in
the Rosensweig Instability}
\author{Adrian Lange}
\institute{
Universit\"at Magdeburg, Institut f\"ur Theoretische
Physik, Universit\"atsplatz 2, D-39106 Magdeburg, Germany
}
\pacs{47.20.Ma}{Hydrodynamic stability and instability}
\pacs{75.50.Mm}{Magnetic liquids}

\begin{document}

\maketitle

\begin{abstract}
The dependence of the maximal growth rate of the modes of the Rosensweig
instability on the properties of the magnetic fluid
and the external magnetic induction is studied. An expansion and a fit
procedure are applied in the appropriate ranges of the supercritical induction
$\hat B$. With increasing $\hat B$ the scaling of the maximal growth rate
changes from linear to a combination of linear and square-root-like  scaling.
The scaling of the corresponding wave number alternates from quadratic to
primarily linear. For very small $\hat B$ the dependence of the maximal growth
rate on the viscosity is given. Suggestions are made for experiments to test
the predicted scaling behaviours.
\end{abstract}

\section{Introduction}
The investigation of instabilities in magnetic fluids has a long history where
the most prominent instability being the normal field or Rosensweig instability 
\cite{cowley,rosensweig}. Above a threshold of the induction, 
the initially flat surface exhibits a stationary array of peaks.
Despite its long history, some aspects of the Rosensweig instability
have been addressed only recently: the hexagon-square transition
\cite{abou00} or the wave number selection problem \cite{abou00,lange00}.
The wave number is the absolute value of the wave vector,
$q=|{\bf q}|$, which characterizes small disturbances. The ground
state of a pattern forming system is subjected to such small
disturbances in order to study its stability.
The corresponding growth rate $\omega$ may depend on system
parameters $\{ {\cal P}\}$, e.g. the viscosity of the fluid, and control parameters
$\{ {\cal R}\}$, e.g. an external magnetic field, $\omega =\omega (q; \{ {\cal P}\},
\{ {\cal R}\})$. Generally it is assumed that in the linear stage of the pattern
forming process the wave number with the largest growth rate will prevail. 
Therefore this mode is called linearly most unstable mode. Due to its role
in the pattern formation it is of particular interest to examine the maximal
growth rate $\omega_m$, and its dependence on the different parameters. 

This dependence has received only limited attention in classical hydrodynamic
systems. For the K\"uppers-Lortz (KL) instability in Rayleigh-B\'enard
convection rotated about a vertical axis, the growth rate of different
KL angles were calculated for one fixed rotation rate and different temperature
differences \cite{hu98}. In surface-tension-driven B\'enard convection
the growth rates are calculated for two fixed values of heat loss and different
temperature differences \cite{thess95}. But in both systems it was not analysed
how $\omega_m$ depends on the control parameters.
In the problem of convection for autocatalytic reaction fronts, the maximal
growth rate was analysed only for the two distinct limits of infinite and
zero thermal diffusivity \cite{edwards91}. In both cases a power law was found
relating the increase of the maximal growth rate with the density
difference in the fluid.
For the Rayleigh-Taylor instability of superposed incompressible viscous
fluids, the maximal growth rate was analysed for appropriately scaled
densities and viscosities in \cite{menikoff77}. The dependence of
$\omega_m$ on the remaining parameter, the scaled surface tension,
is given by a single function covering the whole range of possible
values of the surface tension. With respect to experiments, no
measurements focusing on the maximal growth rate have yet been undertaken.

Particularly in magnetically or electrically driven systems the fastest growing
mode is experimentally accessible. If the field is increased with a sudden
jump, a pattern characterized by the linearly most unstable mode should be observed.
The validity of this conclusion was shown recently for the Rosensweig instability
in magnetic fluids \cite{lange00}. The lack of studies and the experimental
access motivates this letter, where the dependence of the maximal
growth rate on the properties of the magnetic fluid and the external
magnetic induction is studied.

\section{System}
A horizontally unbounded layer of an incompressible, nonconducting, and viscous
magnetic fluid of thickness $h$ and constant density $\rho$ is considered. The
fluid is bounded from below by the bottom of a container made of a magnetically
impermeable material and has a free surface with air above. The electrically
insulating fluid justifies the stationary form of the Maxwell equations, which
reduce to the Laplace equation for the magnetic potentials in the region of
the  container, the magnetic fluid, and the air. It is assumed that the magnetization
of the magnetic fluid depends linearly on the applied magnetic field,
${\bf M} =(\mu_r -1){\bf H}$, where $\mu_r$ is the relative permeability
of the fluid.

In a linear stability analysis, all small disturbances from the basic
state are decomposed into normal modes, i.e., into components of the form
$\exp [-i(\omega\,t -{\bf q}\,{\bf r})]$ with ${\bf r} =(x,y)$.
If ${\rm Im}(\omega) >0$, initially
small undulations will grow exponentially and the originally horizontal surface
is unstable. Therefore $\omega$ is commonly called the growth
rate, which is in fact true only for its imaginary part in the chosen normal
mode ansatz. The linear stability analysis leads to the dispersion relation
\cite{abou97,weilepp,mueller98}
\begin{eqnarray}
  \nonumber
  0 & = &{\nu^2\over \tilde q\coth(\tilde q h)- q\coth (qh)}\biggr\{
    \tilde q\left[ 4q^4 +(q^2+\tilde q^2)^2\right]\coth(\tilde q h)
    -q\left[ 4q^2\tilde q^2+(q^2 +\tilde q^2)^2\right]\tanh (qh)\\
  &&-{4q^2\tilde q (q^2+\tilde q^2)\over \cosh (qh) \sinh (\tilde q
    h)}\biggr\}
    +\tanh (qh) \left[ gq +{\sigma\over \rho}q^3-{\mu_0 \mu_r M^2\over \rho}
   \Lambda (qh)\,q^2\right]\, ,
  \label{eq:1}
\end{eqnarray}
where $\nu$ is the kinematic viscosity, $\sigma$ the surface tension of the
magnetic fluid with air, $M$ the absolute value of the magnetization,
${\bf g}=(0,0,-g)$ the acceleration due to gravity, 
$\mu_0$ the permeability of free space, $\tilde q =\sqrt{q^2
-i\omega/\nu}$, and
$\Lambda (qh) = [{\rm e}^{qh} (1+\mu_r) + {\rm e}^{-qh}(1-\mu_r)]/
[{\rm e}^{qh}(1+\mu_r)^2 - {\rm e}^{-qh}(1-\mu_r)^2]$.
For not too thin layers the approximation with an infinitely thick layer
is applicable \cite{lange00,abou00}. Therefore the starting point
is the dispersion relation (\ref{eq:1}) for $h\,\rightarrow\,\infty$
\cite{salin93}
\begin{equation}
  \label{eq:2}
  \left( 1-{i\omega\over 2\nu q^2}\right)^2 +
  {1\over 4\rho \nu^2 q^4}\left[ \rho g q + \sigma q^3 - {(\mu_r -1)^2\over
  (\mu_r +1)
  \mu_0 \mu_r} B^2 q^2\right] = \sqrt{ 1 -{i\omega \over \nu q^2} }\; .
\end{equation}
Dimensionless quantities are introduced for all lengths, the induction, the
time, and the viscosity,
\begin{eqnarray}
  \label{eq:3}
  \bar l &=& q_c\; l ,\hskip 3.15 cm \bar B = {B\over B_{c,\infty}} ,\\
  \label{eq:4}
  \bar t &=& {g^{3/4}\rho^{1/4}\over \sigma^{1/4}}\; t ={t\over t_c} ,
  \qquad\qquad \bar \nu ={ g^{1/4} \rho^{3/4} \over \sigma^{3/4} }\; \nu\, ,
\end{eqnarray}
where $t_c$ is the so-called capillary time. For $\bar\omega =i\bar\omega_2$
with $\bar\omega_2 >0$, the real part of eq.~(\ref{eq:2}) reduces to
\begin{equation}
  \label{eq:5}
  f_+ (\bar q , \bar\omega_2 ;\bar\nu ,\bar B):=
  \left(\bar\nu + {\bar\omega_2 \over 2\bar q\, ^2}\right)^2
  +{\bar q +\bar q\, ^3 -2 \bar B^2 \bar q\, ^2\over 4 \bar q\, ^4}
  -\bar\nu^2\sqrt{ 1 +{\bar\omega_2 \over \bar\nu\bar
  q\, ^2} } =0\; .
\end{equation}
The imaginary part of eq.~(\ref{eq:2}) is identically zero. The wave number of
maximal growth, $\bar q_m$, is defined by $\partial\bar\omega_2 /\partial \bar q =0$.
Since $\bar\omega_2$ is given implicitly by $f_+$, the cross section of
$f_+=0$ and $\partial f_+ /\partial \bar q =0$ determines the maximal growth rate
and its corresponding wave number.

\section{Analysis and Results}
An expansion of $\bar B$, $\bar q$, and $\bar \omega_2$ in the following form
\begin{equation}
  \label{eq:6}
  \bar B = 1+\hat B\qquad\qquad
  \bar q = 1+\hat q_m\qquad\qquad
  \bar \omega_2 = 0 + \hat\omega_{2,m}
\end{equation}
leads to an analytical expression of the dependence of $\hat\omega_{2,m}$ and $\hat q_m$
on the induction and the viscosity. All hatted quantities in (\ref{eq:6}) are small,
$(\hat B, \hat q_m, \hat\omega_{2,m})\ll 1$, and
denote the scaled distances from the critical values at the onset of the instability. If
$\bar\nu \gg \hat\omega_{2,m}$, the expansion of $f_+=0$ and its derivative results in
\begin{eqnarray}
  \label{eq:7}
    4\bar\nu\hat\omega_{2,m} - 8\hat B + 3\hat\omega_{2,m}^2
    + 8\bar\nu\hat q_m\hat\omega_{2,m} - 16\hat q_m\hat B  - 4\hat B^2 
    -{\hat\omega_{2,m}^3\over 2\bar\nu}&=&0\\
  \label{eq:8}
    -16\hat B - 8\hat B^2 +4\hat q_m -16\hat B\hat q_m +8\bar\nu\hat\omega_{2,m}
    +8\bar\nu\hat q_m\hat\omega_{2,m} +{\hat\omega_{2,m}^3\over \bar\nu}&=&0\, .
\end{eqnarray}
Considering linear terms in eqs.~(\ref{eq:7}) and (\ref{eq:8}) only, one obtains
$\hat\omega_{2,m}=(2/ \bar\nu)\hat B$ and $\hat q_m =0$. The latter equation describes
an incorrect dependence of the wave number of maximal growth on the applied induction.
The result $\hat q_m =0$ and $\bar q =1$ $\bigr($see eq.~(\ref{eq:6})$\bigr)$,
respectively, implies that the wave number of maximal growth is constant and equal $1$
with increasing induction. Such  a conclusion is true only in the case of infinitely
viscous magnetic fluids as shown in \cite{lange00}. Now higher order terms
of the applied induction are included by the ansatz
\begin{eqnarray}
  \label{eq:9}
  \hat\omega_{2,m} &=& \alpha\hat B+\beta\hat B^2+\gamma\hat B^3 +O(\hat B^4)\; ,\\
  \label{eq:10}
  \hat q_m &=& \delta\hat B^2 +\epsilon\hat B^3 +O(\hat B^4)\; .
\end{eqnarray}
Using such an ansatz, the equations (\ref{eq:7}) and (\ref{eq:8}) contain all
terms that contribute up to third order in $\hat B$. By determining the
expansion coefficients $\alpha$, $\ldots$, $\epsilon$,
the dependence on the parameters viscosity and induction is now given by
\begin{eqnarray}
  \label{eq:11}
    \hat\omega_{2,m} &=& {2\over \bar\nu}\hat B +\left( {1\over \bar\nu}
    - {3\over \bar\nu^3}\right)\hat B^2 +\left({10\over \bar\nu^5}-{3\over
    \bar\nu^3}\right)\hat B^3 + O(\hat B^4)
    \hskip 0.5cm  {\rm for~~} 0\leq\hat B< \bar\nu^2/6\\
  \label{eq:12}
    \hat q_m &=& {6\over \bar\nu^2}\hat B^2 +\left({6\over \bar\nu^2}-{22\over
    \bar\nu^4}\right)\hat B^3 + O(\hat B^4)
    \hskip 2.8cm {\rm for~~} 0\leq\hat B< \bar\nu^2/6\, .
\end{eqnarray}
For scaled inductions larger than $\bar\nu^2/6$, one has to solve the
full implicit equation (\ref{eq:1}) and its derivative with
respect to $q$ numerically. Through the implicit character of
both equations, parameter fits are possible only 
for the dependence of $\hat\omega_{2,m}$ and $\hat q_m$ on $\hat B$. The
fit for an excellent agreement with the numerical data
includes a linear term and a square-root term with respect to $\hat B$,
where the coefficients depend on the used magnetic fluid. Combining both
ranges of $\hat B$, we finally have
\begin{eqnarray}
\nonumber
   \hskip 1.0 cm
   \hat\omega_{2,m} &=& \left\{ \begin{array}{c@{\quad }c}
                         \label{eq:13}
                         {\displaystyle{{2\over \bar\nu}\hat B +\left( {1\over \bar\nu}
                         -{3\over \bar\nu^3}\right)\hat B^2 +\left({10\over \bar\nu^5}-{3\over
                         \bar\nu^3}\right)\hat B^3 }} & 
                         \hskip 0.4cm {\rm for~~} 0\leq\hat B< \bar\nu^2/6
                         \hskip 1.2cm (13)\\
                         c_1\sqrt{\hat B} + c_2\hat B  & 
                         \hskip 0.425cm {\rm for~~} \bar\nu^2/6\ll\hat B\leq 0.4\hskip 0.875 cm (14)
                      \end{array} \right. \\
\nonumber
   \hat q_m &=& \left\{ \begin{array}{c@{\quad }c}
                         {\displaystyle{{6\over \bar\nu^2}\hat B^2
                         +\left({6\over \bar\nu^2} -{22\over
                         \bar\nu^4}\right)\hat B^3 }} & 
                         \hskip 2.65cm {\rm for~~} 0\leq\hat B< \bar\nu^2/6 \hskip 1.25 cm (15)\\
                         c_3\hat B + c_4\sqrt{\hat B}  & 
                         \hskip 2.7cm {\rm for~~} \bar\nu^2/6\ll\hat B\leq 0.4\, , \hskip 0.7 cm (16)
                      \end{array} \right. 
\end{eqnarray}
where the four coefficients $c_i$, $i=1, \ldots , 4$, are given in table~\ref{table:1} for
eleven different fluids. These magnetic fluids are made of magnetite nanoparticles
dispersed in a carrier liquid (synthetic ester for APG fluids,
light petrol for EMG fluids). The nanoparticles are coated with a
layer of chemically adsorbed surfactants to avoid agglomeration.

\begin{table}
\caption{Material \protect\cite{ferrofluidics} and fit parameter of the magnetic fluids.}
\label{table:1}
\begin{center}
\begin{tabular}{ccccccccc}
Fluid    & $\rho$ (kg m$^{-3}$) & $\nu$ (m$^2$ s$^{-1}$) & $\sigma$ (kg s$^{-2}$) & $\bar\nu^2/6$  &  $c_1$  & $c_2$  & $c_3$  &  $c_4$  \\ 
EMG 909  & $1.02\cdot 10^3$ & $5.880\cdot 10^{-6}$ & $2.65\cdot 10^{-2}$ & $1.4\cdot 10^{-4}$ & 1.03 & 3.90 & 3.46 & -0.07 \\
EMG 901  & $1.53\cdot 10^3$ & $6.540\cdot 10^{-6}$ & $2.27\cdot 10^{-2}$ & $3.9\cdot 10^{-4}$ & 1.16 & 2.95 & 3.32 & -0.10 \\
APG J16  & $1.01\cdot 10^3$ & $2.475\cdot 10^{-5}$ & $3.30\cdot 10^{-2}$ & $1.7\cdot 10^{-3}$ & 0.72 & 3.52 & 3.12 & -0.20 \\
APG J14  & $1.06\cdot 10^3$ & $2.830\cdot 10^{-5}$ & $3.40\cdot 10^{-2}$ & $2.3\cdot 10^{-3}$ & 0.69 & 3.44 & 3.08 & -0.20 \\
APG S20  & $1.05\cdot 10^3$ & $3.333\cdot 10^{-5}$ & $3.30\cdot 10^{-2}$ & $3.3\cdot 10^{-3}$ & 0.74 & 2.91 & 2.99 & -0.22 \\
APG J12  & $1.11\cdot 10^3$ & $3.604\cdot 10^{-5}$ & $3.40\cdot 10^{-2}$ & $4.0\cdot 10^{-3}$ & 0.58 & 3.29 & 2.95 & -0.23 \\
APG L17  & $1.05\cdot 10^3$ & $5.714\cdot 10^{-5}$ & $3.40\cdot 10^{-2}$ & $9.2\cdot 10^{-3}$ & 0.43 & 3.02 & 2.71 & -0.26 \\
APG J10  & $1.16\cdot 10^3$ & $6.034\cdot 10^{-5}$ & $3.40\cdot 10^{-2}$ & $1.2\cdot 10^{-3}$ & 0.38 & 2.92 & 2.63 & -0.27 \\
APG S11n & $1.15\cdot 10^3$ & $6.957\cdot 10^{-5}$ & $3.30\cdot 10^{-2}$ & $1.6\cdot 10^{-3}$ & 0.33 & 2.80 & 2.51 & -0.28 \\
APG 077n & $1.19\cdot 10^3$ & $9.244\cdot 10^{-5}$ & $3.50\cdot 10^{-2}$ & $2.8\cdot 10^{-3}$ & 0.26 & 2.55 & 2.34 & -0.32 \\
APG 037  & $0.92\cdot 10^3$ & $1.196\cdot 10^{-4}$ & $3.40\cdot 10^{-2}$ & $3.3\cdot 10^{-2}$ & 0.25 & 2.43 & 2.24 & -0.32 \\
\end{tabular}
\end{center}
\end{table}

\begin{figure}
\begin{center}
\includegraphics[width=6.5cm, height=5.6cm, angle=0]
{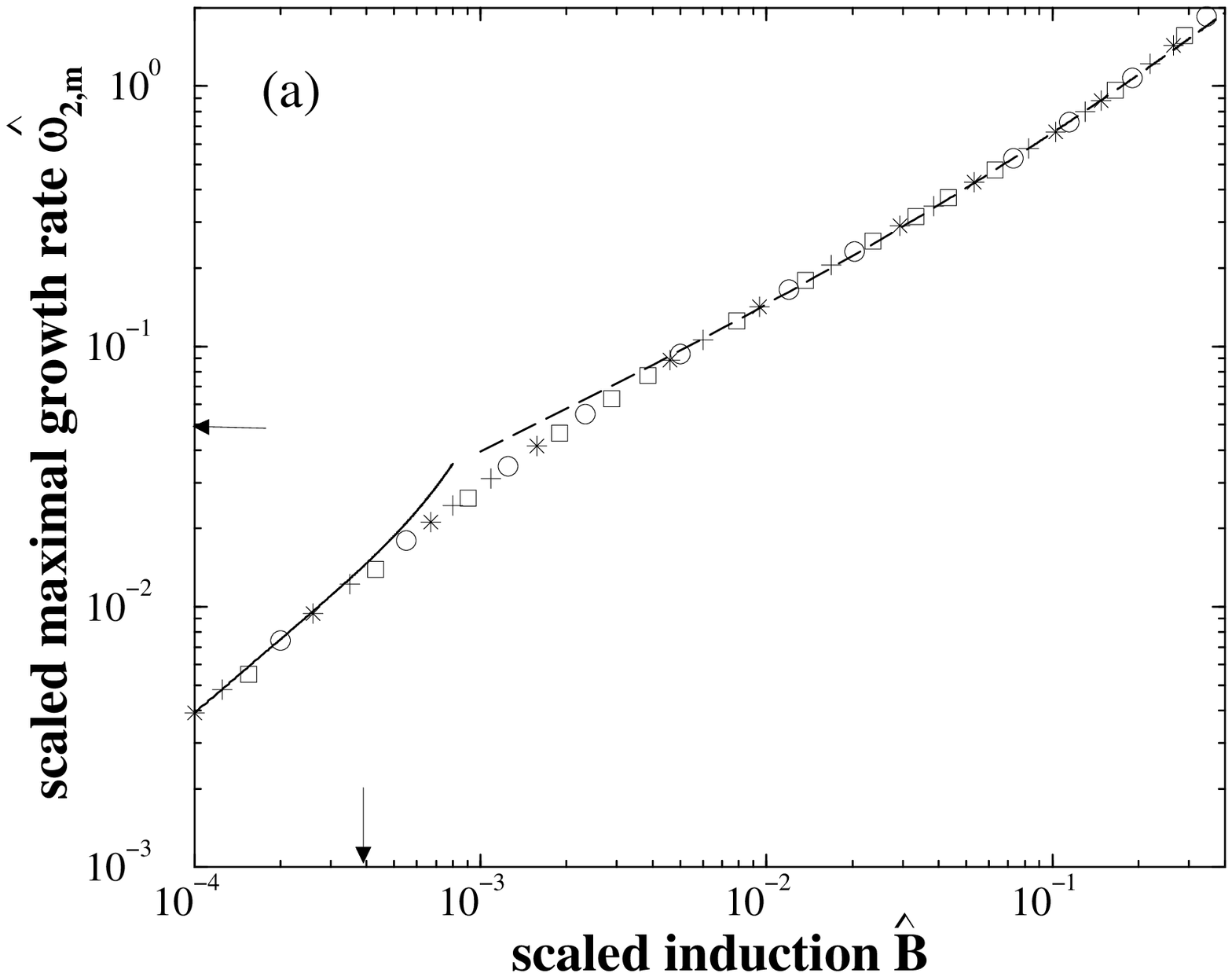}
\hskip 0.8cm
\includegraphics[width=6.5cm, height=5.6cm, angle=0]
{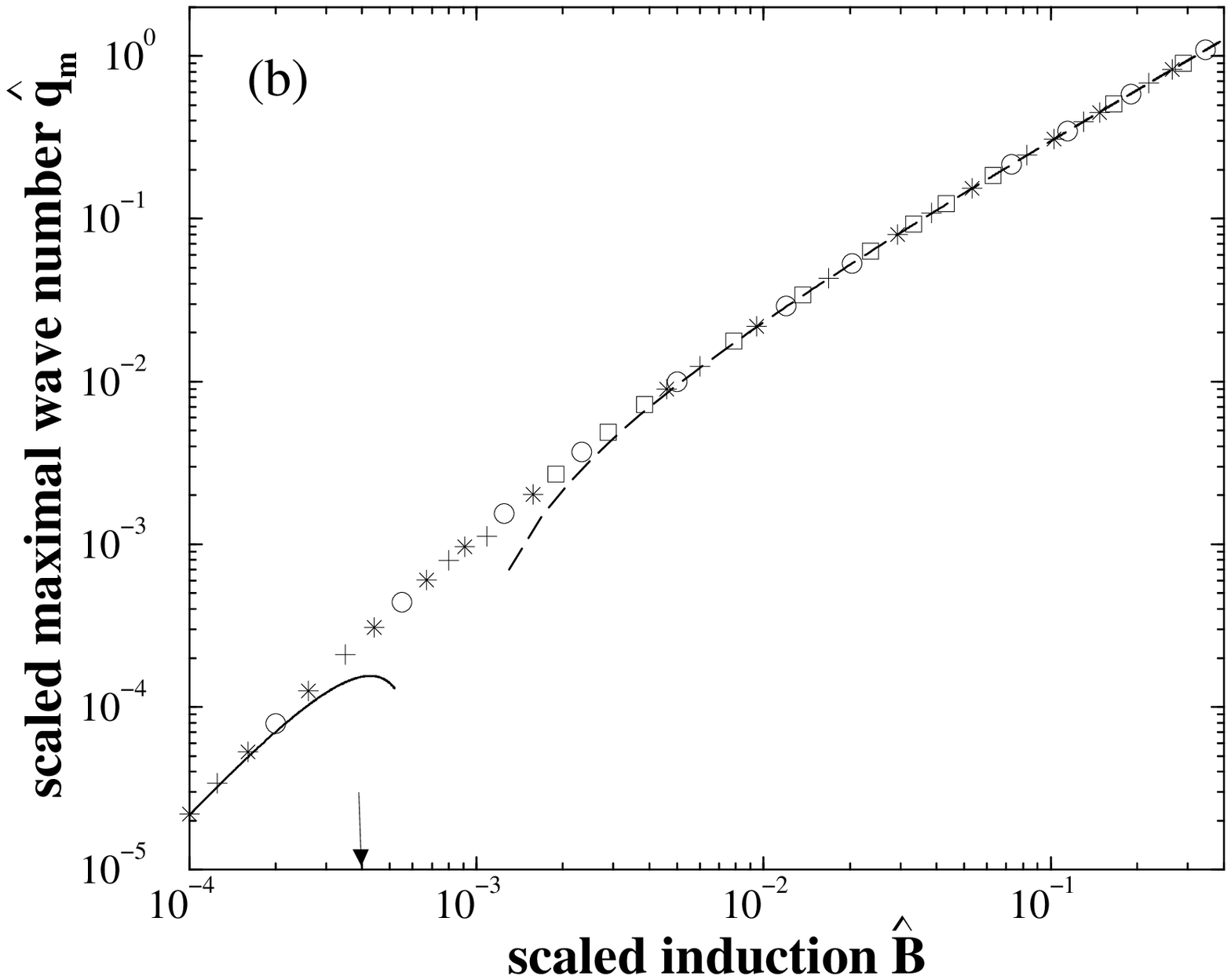}
\caption{Scaled maximal growth rate $\hat \omega_{2,m}$ (a) and scaled maximal
wave number $\hat q_m$ (b) as a function of the scaled supercritical
induction $\hat B$ for EMG 901. The solid lines denote the analytical results
(\ref{eq:11}) for $\hat \omega_{2,m}$ and (13) for $\hat q_m$ in (a) and (b).
The long-dashed lines denote the fits (12) for $\hat \omega_{2,m}$ and (14) for
$\hat q_m$ in (a) and (b). Both scaling regimes are clearly separated.
The data are calculated for $h=100$ mm
($\circ$), $50$ mm ($\ast$), $10$ mm ($+$), and $4$ mm ($\scriptstyle{\square}$)
from eq.~(\protect\ref{eq:1}). The vertical arrows in (a) and (b) indicate
$\bar\nu^2/6$, the horizontal arrow in (a) the dimensionless viscosity $\bar\nu$.
}
\label{fig:1}
\end{center}
\end{figure}

\begin{figure}
\begin{center}
\includegraphics[width=6.5cm, height=5.6cm, angle=0]
{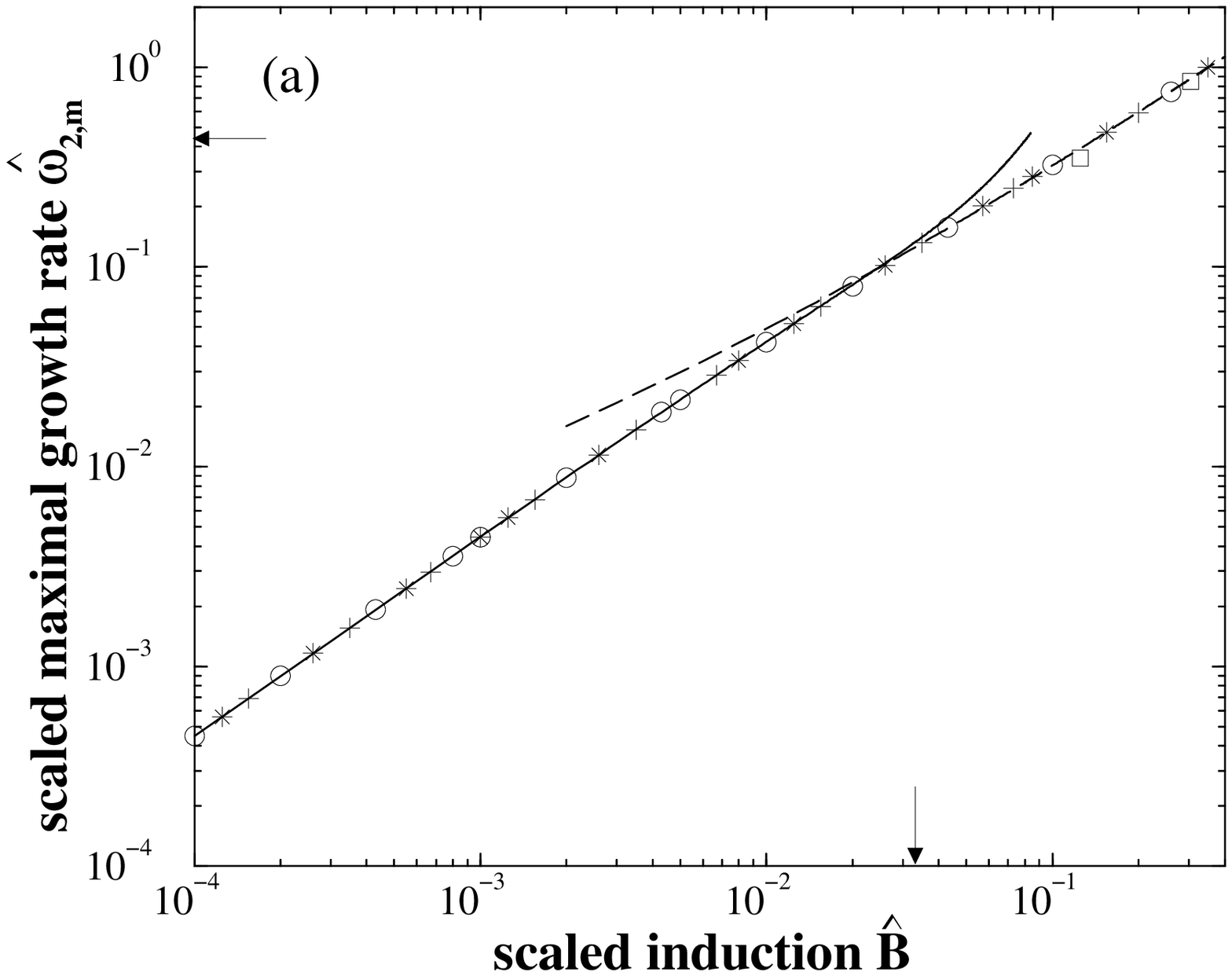}
\hskip 0.8cm
\includegraphics[width=6.5cm, height=5.6cm, angle=0]
{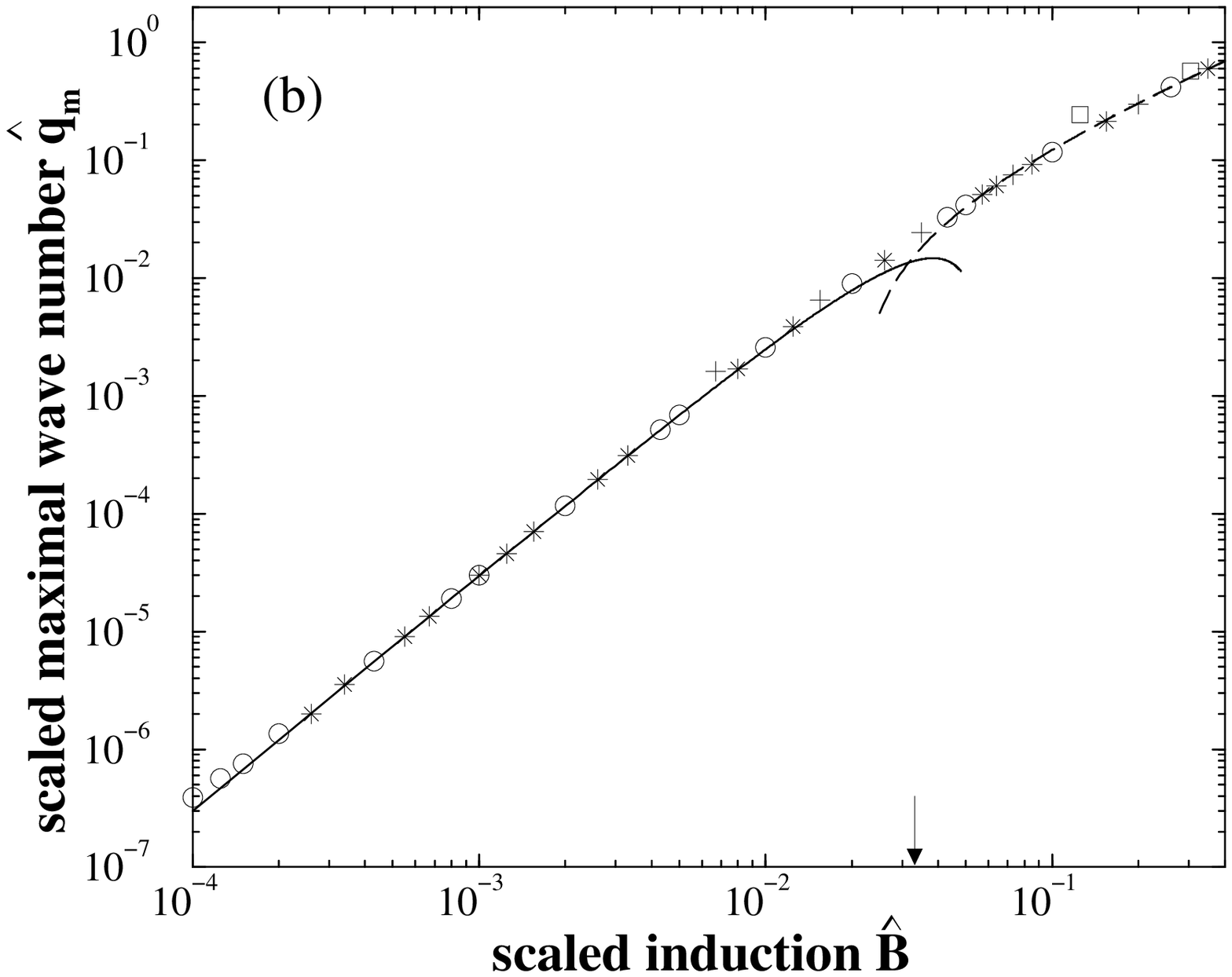}
\caption{Scaled maximal growth rate $\hat \omega_{2,m}$ (a) and scaled maximal
wave number $\hat q_m$ (b) as a function of the scaled supercritical
induction $\hat B$ for APG 037. The transition range between the different
scaling regimes shifts to even higher values of $\hat B$.
Both scaling regimes overlap around $\hat B\sim
0.03$. The symbols, the arrows and the line types are used as in
fig.~\protect\ref{fig:1}.
}
\label{fig:3}
\end{center}
\end{figure}

From eqs.~(\ref{eq:13}-16) and the figures~\ref{fig:1}-\ref{fig:3}
it is evidently that there are two different scaling regimes for the maximal growth
rate as well as for the corresponding wave number.
In all three examples, the assumption for the expansion, $\bar\nu\gg \hat\omega_{2,m}$,
is well accomplished, see horizontal arrows and solid lines in
figs.~\ref{fig:1}(a)-\ref{fig:3}(a).
A lower bound for the transition between the two regimes is given by $\bar\nu^2/6$.
This value is denoted by the vertical arrows in figs.~\ref{fig:1}-\ref{fig:3}.
$\bar\nu$ is a combination of the density, the viscosity, and the surface tension of
the magnetic fluid, see eq.~(\ref{eq:4}). Therefore the transition range
can be matched to experimentally manageable step sizes for the jumplike increase
of the magnetic induction by choosing a proper
fluid. For the chosen three fluids, the transition range shrinks from a broad interval
of $0.0005\leq \hat B_{trans}\leq 0.004$ for EMG 901 (fig.~\ref{fig:1}) to
$\hat B_{trans}\sim 0.03$ for APG 037 (fig.~\ref{fig:3}). This shift is mainly caused
by an increase in the viscosity (see table~\ref{table:1}). As a consequence,
the range of validity of the two-parameter fit in eqs.~(14) and (16) can be replaced
by $\bar\nu^2/6\leq\hat B\leq 0.4$ for the fluid APG 037 (see fig.~\ref{fig:3}).
The appearance of different scaling regimes in the Rosensweig instability make it more
attractive than instabilities with a single scaling such as the Rayleigh-Taylor
instability \cite{menikoff77}.

The maximal growth rate starts to increase {\em linearly} with $\hat B$. Towards the
upper bound of the expansion region, $\hat B\lesssim\bar\nu^2/6$, corrections
appear due to the quadratic and cubic term in (\ref{eq:13}). For $\bar\nu^2/6\ll \hat B$,
the dependence of the maximal growth rate on the induction is given
by a {\em linear} as well as by a {\em square-root} term.
Both terms are equally important since approximations with only one term result
in inferior fits.

For very small $\hat B$, the wave number with maximal growth rate increases
{\em quadratically} with $\hat B$. Near the upper bound of the expansion region
cubic corrections occur. For $\bar\nu^2/6\ll \hat B$, $\hat q_m$ depends
primarily {\em linearly} on $\hat B$ because $c_3\gg |c_4|$.
This linear dependence was already studied in \cite{lange00} and
a good agreement was found between the measured values and the
theoretical data.

\begin{figure}
\begin{center}
\includegraphics[width=6.0cm, height=5.6cm, angle=0]
{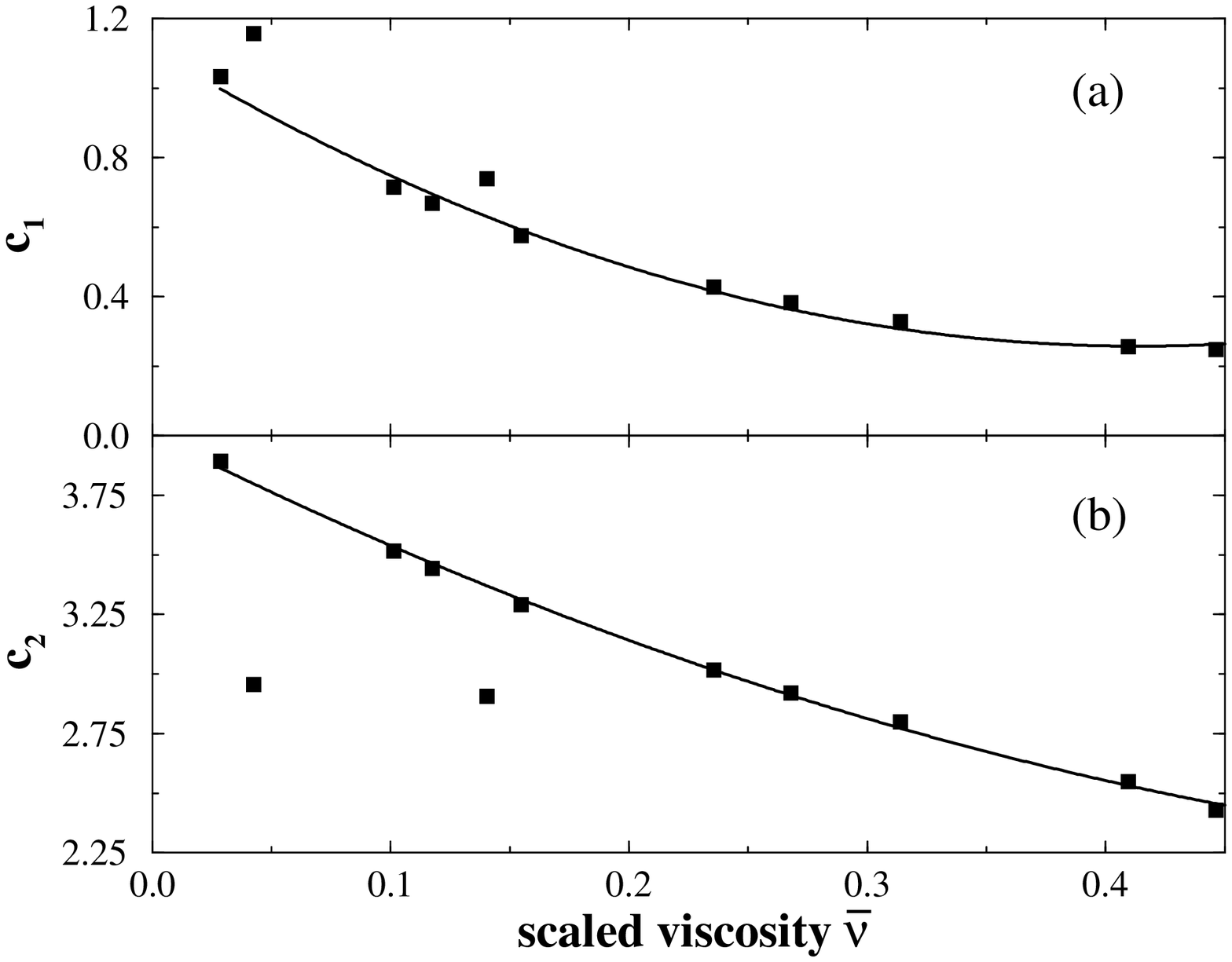}
\hskip 0.5 cm
\includegraphics[width=6.0cm, height=5.6cm, angle=0]
{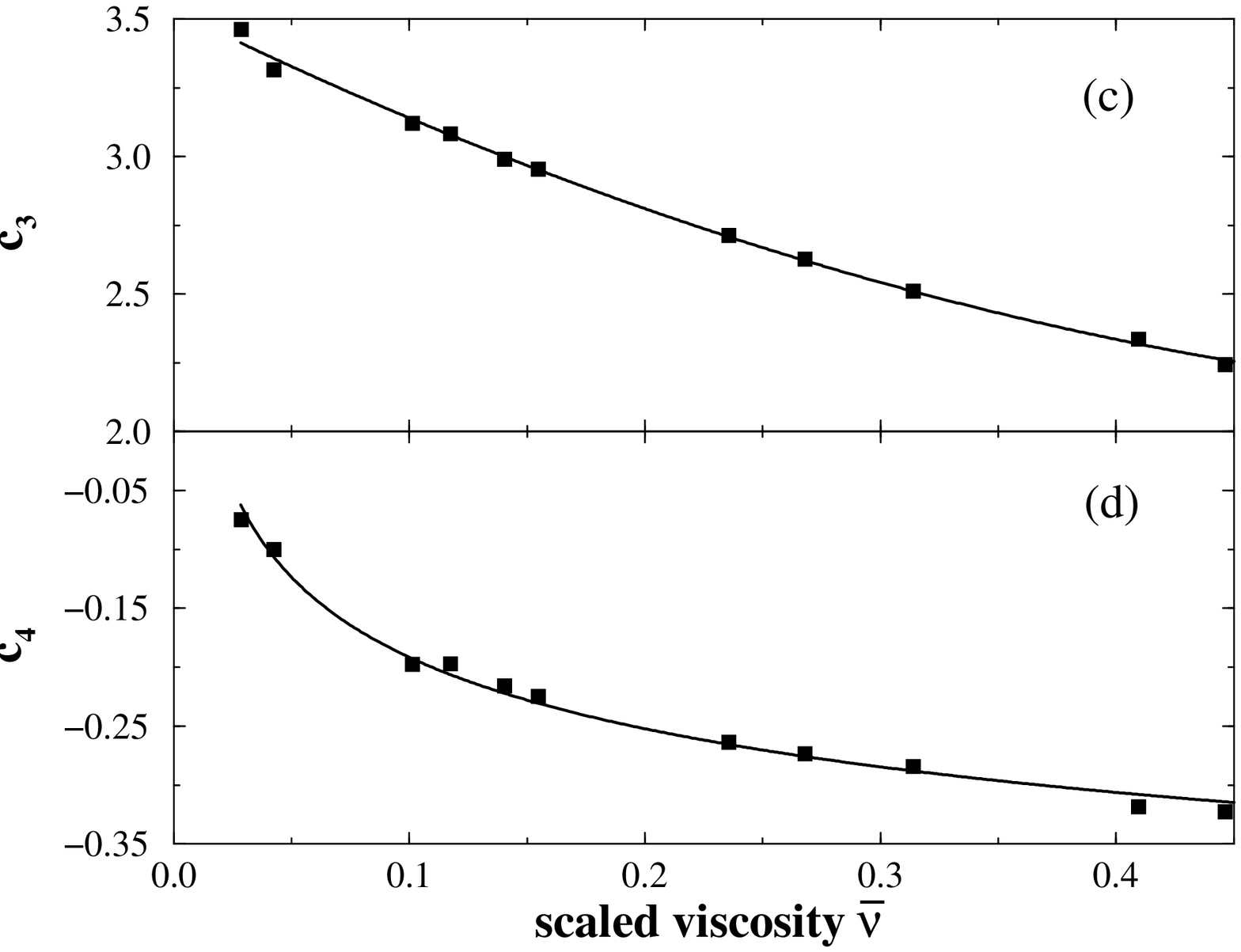}
\hskip 0.8cm
\caption{Dependence of the coefficients $c_1$ (a), $c_2$ (b), $c_3$ (c), and
$c_4$ (d) on the scaled viscosity $\bar\nu$. The values of $c_i$ from
table~\ref{table:1} are plotted as filled squares. The solid lines indicate the
approximations by eqs.~(\ref{eq:15}) and (\ref{eq:16}).}
\label{fig:4}
\end{center}
\end{figure}

The fits~(14) and (16) give only the dependence of $\hat \omega_{2,m}$
and $\hat q_m$ on the scaled magnetic induction. It would beneficial to gain
insight into the dependence on $\bar\nu$, too, at least approximately. Therefore
the set of eleven fluids and their fit coefficients is used to approximate the relations
$c_i=c_i(\bar\nu )$. It is apparent from fig.~\ref{fig:4} that all coefficients
exhibit a smooth dependence on $\bar\nu$ for $0.03\lesssim \bar\nu\lesssim 0.45$.
Only for $\bar\nu\simeq 0.04$ and $\simeq 0.14$, the values of $c_1$ and $c_2$
show deviations. Since $\hat\omega_{2,m} \sim\sigma^{1/4}\omega_{2,m}$,
inaccuracies in the surface tension $\sigma$ may result in incorrect values of $c_1$
and $c_2$. From experiments with surface wave damping on ordinary fluids it is
known that deviations in the value of the surface tension due to contamination
may account for differences
in the order of $20$\% \cite{henderson94,howell00}. Excluding the large
deviations, the following dependencies of $c_i$ on $\bar\nu$ are
suggested (see solid lines in fig.~\ref{fig:4})
\setcounter{equation}{16}
\begin{eqnarray}
\label{eq:15}
c_1 &=& 1.11 - 4.13\bar\nu + 5.0\bar\nu^2 \hskip 0.85 cm
c_2 = 4.0 - 5.0\bar\nu + 3.44\bar\nu^2\; ,\\
\label{eq:16}
c_3 &=& 3.53 -4.21\bar\nu + 3.06\bar\nu^2\qquad
c_4 = -0.75 + 0.38/\sqrt[6]{\bar\nu}\; .
\end{eqnarray}
It is emphasized that the data cover a range of about 2 orders of magnitude
in the kinematic viscosity $\nu$.

Comparing the numerical data and the analytical results for $h\rightarrow\infty$
of $\hat q_m$, one realizes that with increasing fluid viscosity thicker layers
are necessary for a good agreement. This is illustrated at a fixed value of
$\hat B =0.0002$ for the three different fluids. For EMG 901 the analytical
value according to eq.~(15) is already reached at a finite thickness of
$h=7$ mm. For APG S20 the layer thickness has to be $14$ mm before
matching with the value of $\hat q_m$ for $h\rightarrow\infty$. Finally,
for APG 037 a thickness of $h=19$ mm is needed to reach
$\hat q_m$ in the infinitely thick case. Such a sensitivity does not occur for
the maximal growth rate.

\section{Conclusion}
The dependence of the maximal growth rate on the properties of the magnetic fluid
and the external magnetic induction is studied. An expansion and a fit
procedure are applied in the appropriate ranges of the supercritical induction.
Both the maximal growth rate $\hat \omega_{2,m}$ and the corresponding wave number
$\hat q_m$ follow different scaling regimes. Depending on the material parameters
the transition range and its width can be chosen.
The scaling of $\hat \omega_{2,m}$ with respect to $\hat B$ changes from
linear to a combination of linear and square-root-like with increasing $\hat B$.
The scaling of $\hat q_m$ with respect to $\hat B$ alternates from quadratic to
essentially linear with increasing $\hat B$. For very small $\hat B$, $\hat \omega_{2,m}$
depends inversely on the viscosity $\nu$ of the fluid, whereas $\hat q_m$
varies as $\nu^{-2}$ with the viscosity, see eqs.~(\ref{eq:4}, \ref{eq:13}, 15).

\begin{figure}
\begin{center}
\includegraphics[width=6.5cm, height=5.6cm, angle=0]
{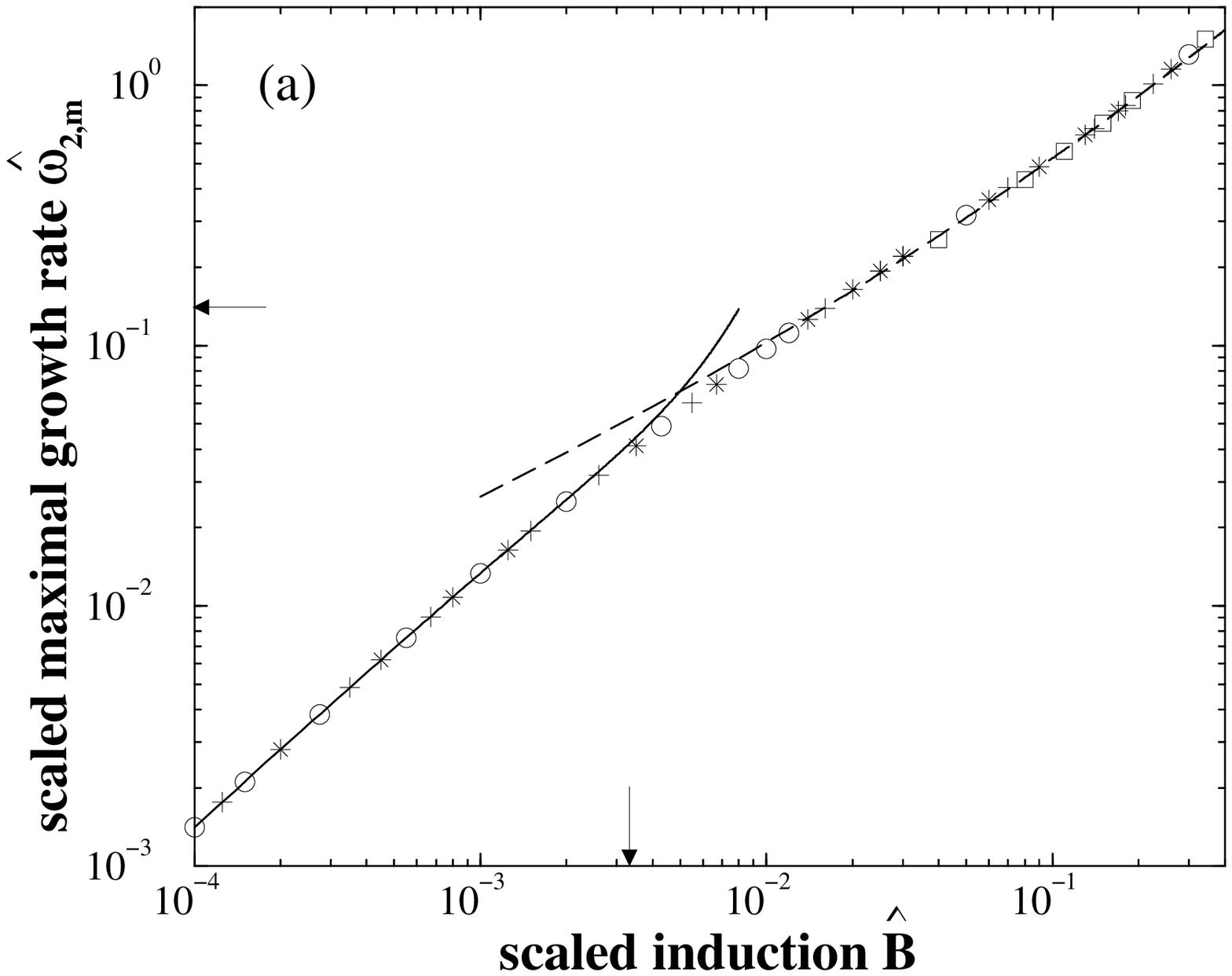}
\hskip 0.8cm
\includegraphics[width=6.5cm, height=5.6cm, angle=0]
{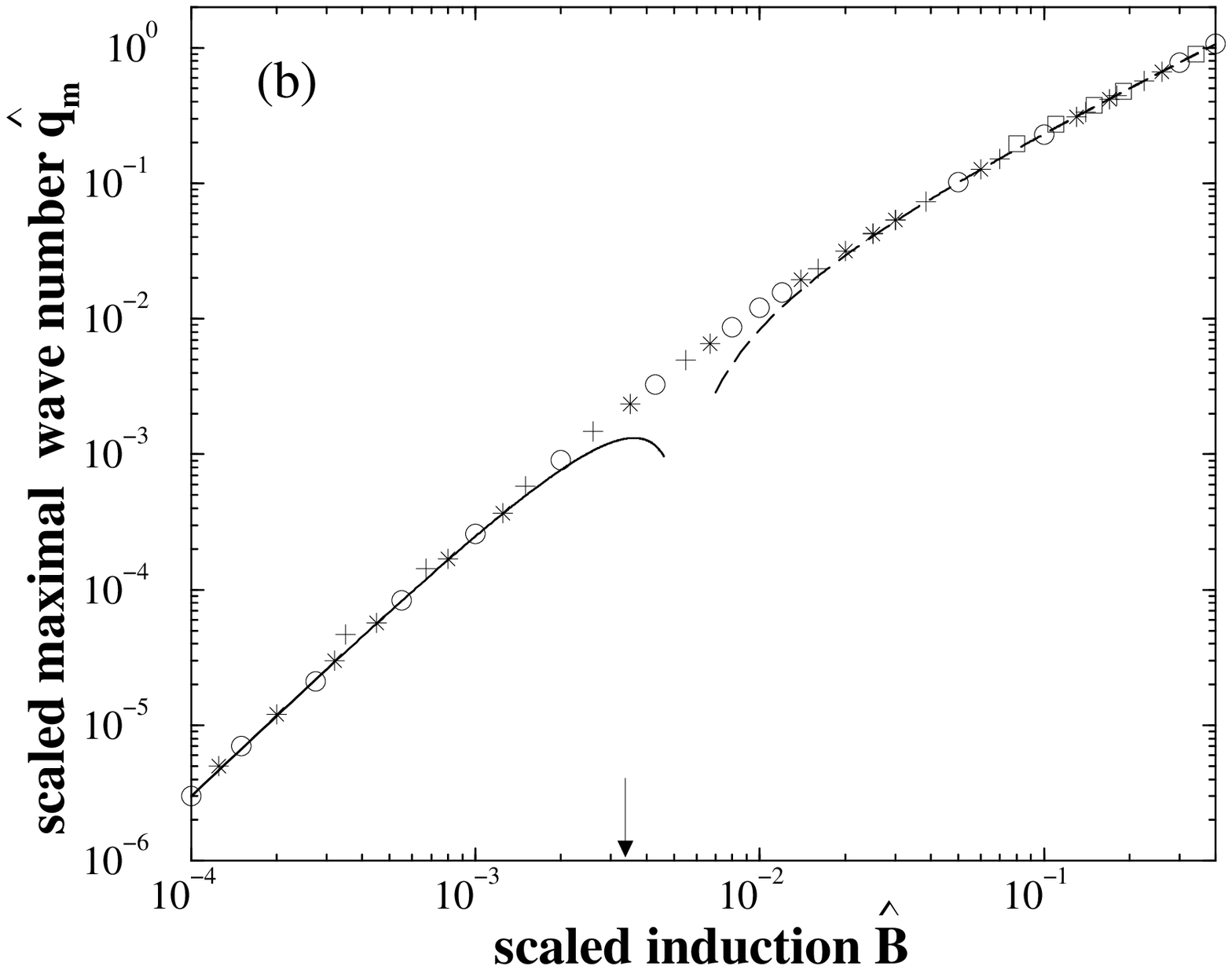}
\caption{Scaled maximal growth rate $\hat \omega_{2,m}$ (a) and scaled maximal
wave number $\hat q_m$ (b) as a function of the scaled supercritical
induction $\hat B$ for APG S20. The transition range between the different
scaling regimes shifts to higher values of $\hat B$ and its width shrinks.
The symbols, the arrows and the line types are used as in fig.~\protect\ref{fig:1}.
}
\label{fig:2}
\end{center}
\end{figure}

The use of magnetic fluids with properties as APG S20 are suggested to
test experimentally the predicted scaling regimes. The range, where
the expansion holds lies within the experimentally realizable step size
for the jumplike increase of the induction. A clear separation between
the different scaling regimes is present (see fig.~\ref{fig:2}). Thus
each regime can be examined independently. Since magnetic fluids
are opaque and feature poor reflectivity, a radioscopic detection
technique is the best method for this purpose. By utilizing the attenuation
of X-rays, the dynamic and static properties of the surface deformation
can be measured with high accuracy \cite{richter01}. Applying
this method to the problem presented here would fill the gap of
experiments measuring especially maximal growth rates.

\acknowledgments
The author thanks B. Reimann and R. Richter for stimulating discussions
and is indebted to B. Huke for bringing into attention calculational
errors. R. Richter is acknowledged for sending a preprint of reference
\cite{richter01} and J. Berg for critical reading of the manuscript.
This work was supported by the Deutsche Forschungsgemeinschaft under
Grant EN 278/2 and LA 1182/2-1.

\begin{thebibliography}{99}
\bibitem{ferrofluidics}
Material data are taken from data sheets of Ferrofluidics Corporation.
\end{thebibliography}
\end{document}